\documentclass[amsmath, amssymb,10pt, aps, prb, twocolumn, notitlepage, eqsecnum, longbibliography, superscriptaddress]{revtex4-1}
\usepackage{graphicx}
\usepackage{enumitem}
\usepackage{calc}
\usepackage{braket}
\usepackage{ulem}   % to strike things out
\usepackage{slashed}
\usepackage{wasysym}
\usepackage{amsthm,amsmath,amsfonts,amssymb,verbatim,color}
\usepackage{lipsum}
\usepackage{bm}
\usepackage{epsfig,slashed}
\usepackage{hyperref}
\usepackage{flafter}
\usepackage{mathtools}
\usepackage{placeins}

\usepackage{float}

\newcommand{\hsigma}{{\hat\sigma}}

\newcommand{\cL}{{\cal L}}

\newcommand{\cO}{{\cal O}}

\DeclareMathOperator{\sign}{sign}

\newcommand{\NuFSSpointtwo}{\ensuremath{ 3.163 \pm 0.041}}

\graphicspath{{data/}}

\begin{document}

\title{Non-Anderson critical scaling of the Thouless conductance in 1D}

\author{Bj\"orn Sbierski}
\affiliation{Department of Physics, University of California, Berkeley, California 94720, USA}

\author{Sergey Syzranov}
\affiliation{Physics Department, University of California, Santa Cruz, California 95064, USA}

\begin{abstract}
	We propose and investigate numerically a one-dimensional model which exhibits a non-Anderson disorder-driven transition. Such transitions have recently been
	attracting a great deal of attention in the context of Weyl semimetals, one-dimensional systems with long-range hopping and high-dimensional
	semiconductors. Our model hosts quasiparticles with the dispersion $\pm |k|^\alpha\sign k$ with $\alpha<1/2$ near two points (nodes) in momentum space
	and includes short-range-correlated random
	potential which allows for scattering between the nodes and near each node.
	In contrast with the previously studied models in dimensions $d<3$, the model considered here exhibits a critical scaling of the Thouless 
	conductance which allows for {an accurate} determination of the critical properties of the non-Anderson transition, with a precision
	significantly exceeding the results obtained from the critical scaling
	of the density of states, usually simulated at such transitions. 
	We find that in the limit of the vanishing parameter $\varepsilon=2\alpha-1$ the correlation-length exponent 
	$\nu=2/(3|\varepsilon|)$ at the transition is inconsistent with the prediction $\nu_{RG}=1/|\varepsilon|$ of the perturbative renormalisation-group analysis. 
	Our results allow for a numerical 
	verification of the convergence of $\varepsilon$-expansions
	for non-Anderson disorder-driven transitions and, in general, interacting field theories near critical dimensions.
\end{abstract}

%%%%%%%%%%%%%%%%%%%%%%%%%%%%%%%%%%%%%%%%%%%%%%%%%%%%%%%%%%%%%%%%%%%%%%%%%%%%%%%%%%%%%%%%%%%%%%

\maketitle

%%%%%%%%%%%%%%%%%%%%%%%%%%%%%%%%%%%%%%%%%%%%%%%%%%%%%%%%%%%%%%%%%%%%%%%%%%%%%%%%%%%%%%%%%%%%%%
\section{\label{sec:Introduction}Introduction}

The desire to identify and understand
 universal critical properties of phase transitions in disordered systems has been
motivating, over several decades, advances in numerical and analytical descriptions of
disordered and interacting systems, such as new field-theoretical approaches~\cite{Efetov:book,Kamenev:book,BelitzKirkpatrick:review}, 
renormalisation-group methods~\cite{DasguptaMa,Fisher:strongRG,AltmanEtAl:OneDRG,Levitov_1990,Levitov2,Burin:MBLclaims} and scaling theories~\cite{Cardy:ScalingBook}.
A significant portion of those developments was driven by the studies of the Anderson localisation-delocalisation
transitions (see Refs.~\onlinecite{Mirlin:review,MirlinEvers:R2,Efetov:book} for a review),
commonly believed to be the only possible disorder-driven transitions in non-interacting systems.

%Despite their fundamental and experimental significance and intense studies,
%accurate analytical description of the critical properties of Anderson transitions
%in spatial dimensions not close to $d=2$ still remains challenging.

The last several years have also seen an upsurge of research activity (see Ref.~\onlinecite{Syzranov:review} for a review)
on non-Anderson disorder-driven transitions, i.e. disorder-driven transitions in universality classes distinct from those of
Anderson localisation. Such transitions (or possibly sharp crossovers~\footnote{\label{foot1} It has recently been debated (for example, in 
	Refs.~\onlinecite{Nandkishore:rare,Sbierski:WeylVector,Mafia:rare,Gurarie:AvoidedCriticality,BuchholdAltland:rare,BuchholdAltland:rare2}),
	in the context of Weyl semimetals,
	if these transitions survive non-perturbative rare-region effects or get converted to sharp crossovers.
	We do not distinguish between true phase transition and sharp crossovers in this paper,
	provided the non-Anderson critical scaling exists in a parametrically
	large interval of lengths.})
have been first proposed~\cite{Fradkin1,Fradkin2}
for Weyl semimetals. They have later been demonstrated~\cite{Syzranov:WeylTransition,SyzranovGurarie:duality,Syzranov:review} to occur in all systems
with the power-law quasiparticle dispersion $\propto k^{\alpha}$ in high
dimensions~\cite{Syzranov:review,SyzranovGurarie:duality} $d>2|\alpha|$.
These systems include, but are not limited to,
arrays of trapped ultracold ions which exhibit power-law hopping of excitations~\cite{Monroe:longrange,Islam:longrange,Blatt:chain1,Blatt:chain2,RodriguezMalyshev:veryveryFirst,Rodriguez:firstPowerLaw,Malyshev:firstPowerLaw,Moura:firstPowerLaw,Xiong2003:roughScaling,Garttner:longrange},
high-dimensional semiconductors, quantum kicked rotors~\cite{Syzranov:review} and certain disordered
 supercoductive systems~\cite{TikhonovFeigleman:PowerLawSupercond}.
Apart from non-Anderson universality classes, such systems display critical scaling of the density of states 
(which does not exist for Anderson transitions), unconventional behaviour of Lifshitz tails~\cite{Suslov:rare,Nandkishore:rare,Syzranov:unconv},
energy-level statistics and ballistic-transport properties~\cite{Syzranov:review}.

Most studies of the unconventional non-Anderson transitions to date have been focussing on Weyl semimetals,
the best known and experimentally available systems predicted to exhibit them. 
Obtaining the critical exponents at these transitions in Weyl semimetals
has been the goal of dozens of analytical and numerical studies (see, e.g., Refs.~\onlinecite{ShindouMurakami,GoswamiChakravarty,RyuNomura,KobayashiOhtsukiHerbut:scaling,Syzranov:WeylTransition,Syzranov:unconv,LouvetFedorenko:theirFirst,RoyJuricic:superuniversality,Syzranov:TwoLoop,PixleyHuse:missedPoint,Sbierski:superAccurate,BeraRoy:inaccurateNumerics,LiuOhtsuki:LateNumerics,Balog2018:porousMedium}).

3D Weyl semimetals do not have any small parameters at the transition point and all analytical methods of their description
are, strictly speaking, uncontrolled. In contrast with the Anderson localisation transitions in 3D, certain features of the non-Anderson transitions
in Weyl semimetals are described remarkably accurately by means of perturbative one-loop RG calculations
controlled by the parameter $\varepsilon=2-d$ with setting $\varepsilon=-1$ at the end of the calculation.
For example, 
the one-loop RG result $z=3/2$ for the dynamical critical exponent $z$ matches the numerical results~\cite{KobayashiOhtsukiHerbut:scaling,PixleyHuse:missedPoint,Sbierski:superAccurate,BeraRoy:inaccurateNumerics,LiuOhtsuki:LateNumerics} 
within the error of the numerical simulations (1-2\%).

Other results, such as the values of the correlation-length exponent $\nu$, are more controversial (see Ref.~\onlinecite{Syzranov:review} for a review): most
numerical studies (see, e.g., Refs.~\onlinecite{KobayashiOhtsukiHerbut:scaling,RoyJuricic:superuniversality,PixleyHuse:missedPoint,Pixley2015a:AndersonLocalization,Sbierski:superAccurate,BeraRoy:inaccurateNumerics,LiuOhtsuki:LateNumerics,Klier2019:SCBAtransition}) report errors as large as $10-15\%$, with up to $50\%$ differences in the values of $\nu$ between different studies.
The errors in determining the 
correlation-length exponents at such transitions come from the inaccuracies
of the simulations of the critical scaling of the density of states, which have been used to determine the critical properties of the transition
in all studies to date, with the exception of Ref.~\onlinecite{Sbierski:superAccurate}
(where a finite-size analysis of the conductance has been carried out). 
Because the density of states vanishes or nearly vanishes on one side of a non-Anderson transition,
accurate determination of the transition point is challenging,
which may lead to substantial errors in determining
the correlation-length exponent. Furthermore, the transition point may be additionally obscured by rare-region effects,
whose role is also being debated in the literature~\cite{Note1}.

These controversies, together with the need for better understanding of non-Anderson criticality,
has motivated us to propose and study in this paper a model exhibiting non-Anderson disorder-driven transitions
which, on the one hand, is controlled by a small tunable parameter $|\varepsilon|\ll1$ and, on the other hand, exhibits the critical scaling 
of a non-vanishing observable allowing for an accurate determination of the critical properties.

We study a one-dimensional disordered chain where the quasiparticle dispersion hosts two nodes, i.e. points
with the chiral dispersion 
$\pm |k|^\alpha \sign k$, where the momentum $k$ is measured from each node and $\alpha<1/2$.
While transitions in one-dimensional models with power-law dispersions have been studied
numerically previously~\cite{RodriguezMalyshev:veryveryFirst,Rodriguez:firstPowerLaw,Malyshev:firstPowerLaw,Moura:firstPowerLaw,Xiong2003:roughScaling,Garttner:longrange},
the model considered here combines several ingredients which allow us to determine the critical properties of the transition 
rather precisely.

On the one hand, the energies of the nodes are robust against disorder (similarly to that in Ref.~\onlinecite{Garttner:longrange}
and in contrast with Refs.~\onlinecite{RodriguezMalyshev:veryveryFirst,Rodriguez:firstPowerLaw,Malyshev:firstPowerLaw,Moura:firstPowerLaw,Xiong2003:roughScaling}), which is essential for the accurate 
identification of the transition point, 
as the non-Anderson disorder-driven transition takes place for states at only one energy. On the other hand, the system
exhibits a critical scaling of the Thouless conductance, %(in contrast with Ref.~\onlinecite{Garttner:longrange})
a quantity which,
unlike the density of states, remains finite at the transition.
%This allows us to carry out finite-size analysis of the critical scaling of an observable at the non-Anderson transition.

The finite-size scaling analysis of the Thouless conductance allows us
to obtain the correlation-length exponent as a function of the parameter $\varepsilon=2\alpha-1$
and compare the result with the prediction of the analytical perturbative-RG descriptions controlled by the small parameter $\varepsilon$.
We find that, contrary to the common expectation, the numerical result for the correlation-length exponent is inconsistent with the predictions of the perturbative RG approaches.

As non-Anderson disorder-driven transitions may be described by deterministic interacting field theories in the 
replica, supersymmetric or Keldysh representations, the approach we develop here may be used more broadly to verify the convergence of 
the $\varepsilon$-expansions in interacting field theories.

%This paper is organised as follows. \tb{I integrated that in the next section II.}
%\tg{Could you move it to here please? The convention is that the structure of the paper is usually overviewed at the end
%of the introduction. At the end of Section II, it looks a little that we provide an outline 
%in the middle of the paper.}

%%%%%%%%%%%%%%%%%%%%%%%%%%%%%%%%%%%%%%%%%%%%%%%%%%%%%%%%%%%%%%%%%%%%%%%%%%%%%%%%%%%%%%%%%%%%%%%%%%%%%%%%%
%%%%%%%%%%%%%%%%%%%%%%%%%%%%%%%%%%%%%%%%%%%%%%%%%%%%%%%%%%%%%%%%%%%%%%%%%%%%%%%%%%%%%%%%%%%%%%%%%%%%%%%%%

\section{\label{sec:Summary}Summary of the results}

In this work, 
we focus on the numerical analysis of the correlation length-exponent $\nu(\varepsilon)$
as a function of the parameter $\varepsilon=2\alpha-1$, where $\alpha$ characterises the dispersion
$\pm k^\alpha\sign k$ near several points (nodes) in momentum space (see Sec.~\ref{sec:Model}
for a full description of the model and Eq.~\eqref{eq:H0} for the dispersion for all momenta).

We use the critical scaling of a quantity similar to the Thouless conductance $g$, introduced in Sec.~\ref{sec:Thouless},
for an accurate determination of the correlation-length exponent for the parameter $|\varepsilon|$
in the interval $|\varepsilon|=0.1\ldots 0.5$.
Studying the critical properties of the transition for $|\varepsilon|\ll 1$ allows us not only to compare the numerical results with the analytical
perturbative
RG predictions, justified in the limit of small $\varepsilon$, but also to neglect possible non-perturbative rare-region effects~\cite{Note1},
exponentially suppressed~\cite{Syzranov:unconv,Syzranov:review} by the parameter $1/|\varepsilon|$.

\begin{figure}[h!]
	\centering
	\includegraphics[width=\linewidth]{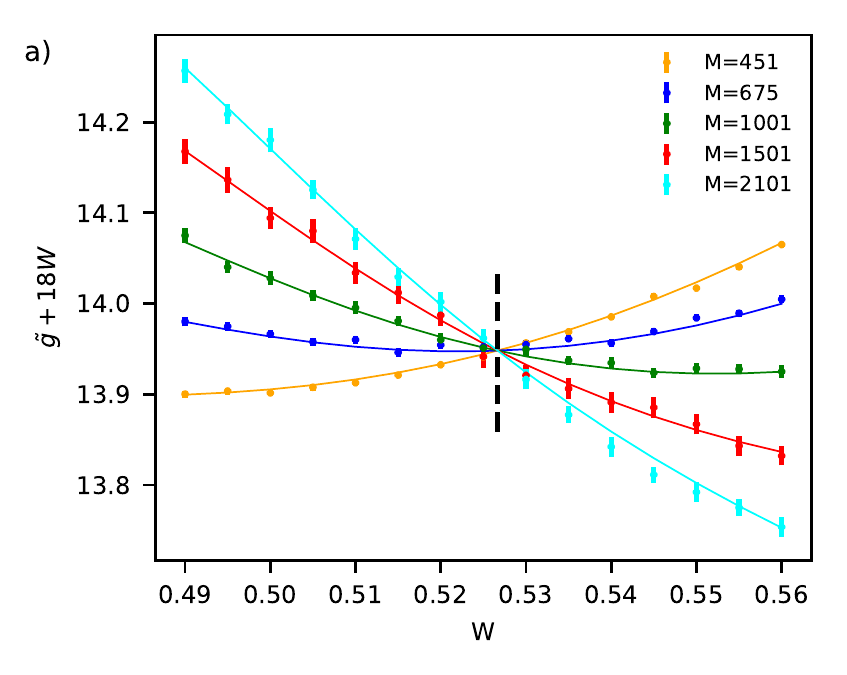}
	\includegraphics{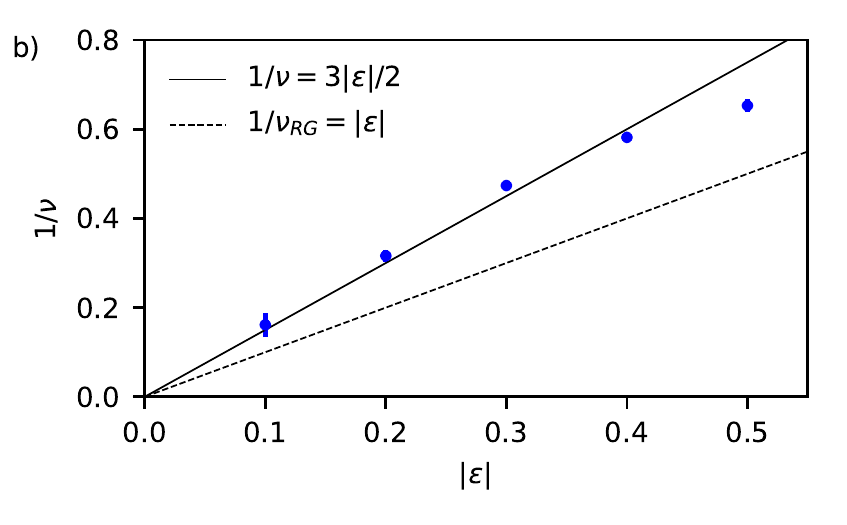}
	\caption{\label{fig:Summary}
		Results for the critical properties of the non-Anderson disorder-driven transition. (a) Median $\tilde{g}$
		of the Thouless conductance distribution for various system sizes $L=2M$ and
		disorder amplitudes $W$. The solid lines show the fourth-order polynomial fit of the scaling function $\tilde{g}\left[u\left(W-W_{c}\right)L^{1/\nu}\right]$ with $u(w)=w+b w^2$ (see Appendix \ref{appendix-FSS}). The vertical line shows the critical amplitude
		$W=W_c$ obtained from the fit. 
		The linear function $18W$ is added to the conductance $\tilde{g}$ on the vertical axis for 
		better visibility of the crossing. 
		(b) The inverse correlation-length exponent $1/\nu$, obtained from the scaling analysis for various $\varepsilon$,
		as a function of $\varepsilon$. The solid line shows the dependence $1/\nu=3|\varepsilon|/2$, which describes 
		accurately
		the numerical data in the limit of small $\varepsilon$, in contrast with the
		prediction $1/\nu_{RG}=|\varepsilon|$ of the analytical RG approach showed by the dashed line.}
\end{figure}

Our results for the critical properties are summarised in Fig~\ref{fig:Summary}. 
Panel (a) shows the finite-size scaling of the median $\tilde{g}$ of the conductance distribution as a function of the system size and disorder strength for representative $\alpha=0.4$
($|\varepsilon|=0.2$). For strong disorder, $\tilde{g}$ decreases with system size $L$, whereas the opposite trend is observed for weak disorder.
These regimes are separated by a critical disorder amplitude $W_c$, a unique crossing point for data traces pertaining to different system sizes. Around $W_c$, we fit the data with a polynomial scaling function (solid lines) to find the correlation-length exponent $\nu$. Based
on the data for five different values of $\varepsilon$, we find the asymptotic
dependence (Fig. \ref{fig:Summary}b)
\begin{equation}
\frac{1}{\nu}=\frac{3}{2}|\varepsilon|\label{eq:y}.
\end{equation}

Surprisingly, this result is in contradiction with the prediction of the perturbative one-loop RG analysis,
\begin{align}
\frac{1}{\nu_{RG}}=|\varepsilon|,
\end{align}
believed to be exact in the limit of $\varepsilon\rightarrow 0$.
We present the details of the analytical RG analysis for the model under consideration in Appendix~\ref{appendix-RG}.

 As further
checks of this unexpected result, we (i) confirm the consistency of our
value of $\nu(|\varepsilon|=0.2)=\NuFSSpointtwo$ with the scaling analysis of the density of states
(see Sec. \ref{sec:DOS} and Fig. \ref{fig:DOS}), and (ii) find an unusual scaling of the critical
dimensionless disorder strength $\gamma_{c}\propto |\varepsilon|^{2/3}$
(Fig. \ref{fig:gammac}) which also contradicts the predictions of the analytical RG approach but is consistent with earlier results for a similar
one-dimensional model.\cite{Moura:firstPowerLaw}

%%%%%%%%%%%%%%%%%%%%%%%%%%%%%%%%%%%%%%%%%%%%%%%%%%%%%%%%%%%%%%%%%%%%%%%%%%%%%%%%%%%%%%%%%%%%%%%%%%%%%%%%%
%%%%%%%%%%%%%%%%%%%%%%%%%%%%%%%%%%%%%%%%%%%%%%%%%%%%%%%%%%%%%%%%%%%%%%%%%%%%%%%%%%%%%%%%%%%%%%%%%%%%%%%%%

\section{\label{sec:Model}Model and Observable}

\begin{figure}[h!]
\centering
\includegraphics{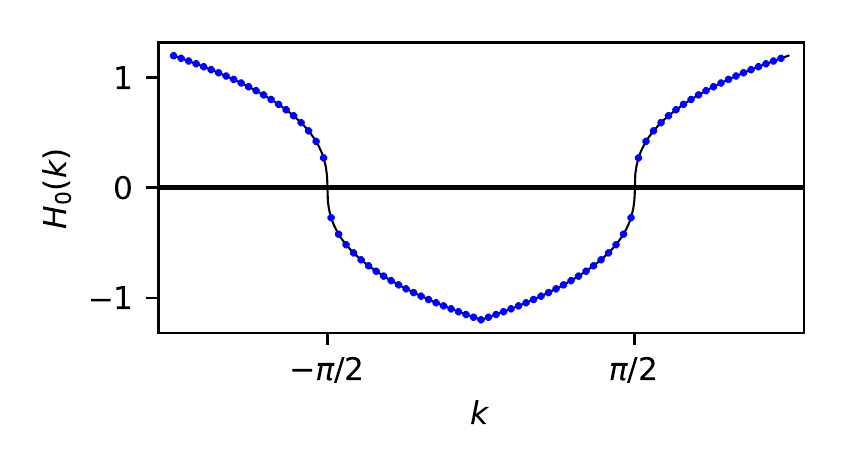}
\caption{\label{fig:ModelObservable}
	The dispersion $H_{0}(k)$ in a disorder-free system as a
function of $k$ for $\alpha=0.4$. The solid line depicts the continuum dispersion
while the blue points indicate the placement of discrete momenta ($M=41$)
in the finite-size version of the model.}
\end{figure}

We consider a model described by the Hamiltonian
\begin{equation}
H_{0}(k)=\begin{cases}
\left|k-\pi/2\right|^{\alpha}\, \sign(k-\pi/2), & 0<k<\pi,\\
\left|k+\pi/2\right|^{\alpha}\,\sign(k+\pi/2), & -\pi<k<0,\\
\infty, & \mathrm{otherwise}.
\end{cases}
\label{eq:H0}
\end{equation}
The dispersion is depicted in Fig. \ref{fig:ModelObservable}. Near the momenta $k=\pm \pi/2$, hereinafter 
referred to as {\it nodes}, the dispersion has the power-law form and lacks reflection symmetry.

The singular dispersion $\propto |k\pm \pi/2|^\alpha\sign (k\pm \pi/2)$ near the nodes corresponds to the power-law hopping
$\propto 1/|x|^{1+\alpha}\sign x$ in coordinate space at large distances $x$. Disorder-driven transitions in similar models with 
power-law hopping, corresponding to the dispersion $\propto k^\alpha$ which is even with respect to the momentum inversion $k\rightarrow -k$), have been studied numerically in Refs.~\onlinecite{RodriguezMalyshev:veryveryFirst,Rodriguez:firstPowerLaw,Malyshev:firstPowerLaw,Moura:firstPowerLaw,Xiong2003:roughScaling}.
Because the dispersion considered here is odd with respect to inverting the momentum measured from the node and the 
momentum states with energies above and below the nodes have equal weights, the energies of the nodes 
are not renormalised if the system is exposed to quenched short-range-correlated disorder with zero average.
Because only the states at the node energies undergo the non-Anderson disorder-driven transitions, the absence of the renormalisation
of the energies of the nodes helps us identify such states more accurately.

A model with the dispersion $\propto k^\alpha \sign k$, the single-node version of the dispersion \eqref{eq:H0},
has been studied in Ref.~\onlinecite{Garttner:longrange}. In that model, the quasiparticles can move only in one direction and cannot
be backscattered. By contrast, the dispersion considered here allows for the scattering between the two nodes
in the presence of the random potential, which
leads to the critical scaling of a quantity similar to the Thouless conductance,~\cite{EdwardsThouless:firstShift,Thouless:review} which we describe below.
The Thouless conductance does not vanish at the critical point, unlike the density of states, 
and thus allows for a significantly more precise determination of the critical properties of the transition.

We discretise the dispersion (\ref{eq:H0}) on a lattice in
momentum space and represent the dispersion in the form $H_{0}=\sum_{k}\left|k\right\rangle H_{0}(k)\left\langle k\right|$
where the sum runs over $k_{m}=\frac{2\pi m}{L}$ with $m=-M,...,M-1$.
Here, $M$ is an odd integer and $L=2M$. These $k$-points
fall symmetrically around the nodes while avoiding the momenta $\pm \pi/2$ of the nodes. 
The density of states in the absence of disorder is given by
$\rho_{\text{clean}}(E)=2\left|E\right|^{1/\alpha-1}/(2\pi\alpha)$,
where the energy $E$ is measured from the energies of the nodes.

We generate disorder on a real-space grid of $4M$ sites with spacing
$1/2$. On each lattice site $i$, the disorder potential $U_{i}$
is drawn from a box distribution $U_{i}\in\left[-\sqrt{3}W,+\sqrt{3}W\right]$
leading to the disorder averaged correlator $\left\langle U_{i}U_{j}\right\rangle _{\text{dis}}=\delta_{i,j}W^{2}$.
In addition, we subtract the mean for each disorder realisation to ensure
$\sum_{i}U_{i}=0$. We then Fourier-transform the generated random potential, $U=\sum_{k\neq k^{\prime}}\left|k^{\prime}\right\rangle U(k^{\prime}-k)\left\langle k\right|$.
Because the potential is real, $U(k)=U^{\star}(-k)$.
We emphasise that, since the dispersion \eqref{eq:H0} is not periodic in momentum space, there is no Umklapp scattering in our model.

The eigenstates of the Hamiltonian $H=H_{0}+U$ are found by exact diagonalisation.
We note that the Hamiltonian matrix in coordinate space is not sparse due to the long-range character of the hopping.

{\it Scaling observable.} %We proceed to discuss the Thouless conductance that is central to
%our scaling analysis. 
In what immediately follows, we describe a quantity similar to the Thouless conductance, which we use in order to identify the non-Anderson disorder-driven
transition. 
The Thouless conductance~\cite{EdwardsThouless:firstShift,Thouless:review, Braun:ThoulessConductance} is
usually defined (up to a coefficient of $1/\delta$, with $\delta$ being the mean level spacing) as 
a response of the energy $E_{n}$ of an eigenstate $\left|n\right\rangle $
to an adiabatic twist of the phase $\phi$ in the boundary conditions~\cite{Thouless:review}.
Such a twist for the system considered here corresponds to the replacement $k\rightarrow k+\frac{2\pi}{L}\phi$
in the dispersion (\ref{eq:H0}).

In order to study the criticality, we analyse the quantity
\begin{equation}
g=\left|{\partial_{\phi}^{2}E_{n_{0}}|_{\phi=0}}/{E_{n_{0}}}\right|,
\label{eq:gTh}
\end{equation}
where $E_{n_{0}}$ is the energy, measured from the node energy, of the eigenstate $n_0$ with the smallest absolute value $|E_{n_{0}}|$
at $\phi=0$.
The Thouless conductance~\cite{EdwardsThouless:firstShift,Thouless:review, Braun:ThoulessConductance}
in systems with a constant density of states in a certain energy interval matches Eq.~\eqref{eq:gTh}
with the replacement $E_{n_{0}} \rightarrow \delta$, where $\delta$ is the mean level spacing in the respective interval.
While the density of states in the system considered here vanishes at the nodes, the quantity $E_{n_{0}}$ in Eq.~\eqref{eq:gTh}
plays a role similar to that of the mean level spacing $\delta$ in the conventional definition of the Thouless conductance.
For simplicity, we refer to the quantity \eqref{eq:gTh} as the Thouless conductance in the rest of the paper.

According to the second-order perturbation theory, a phase twist of $\phi$ at the boundary of the system is 
equivalent to adding the perturbation $H_{\phi}(k)\simeq\left(\frac{2\pi}{L}\phi\right)h_{\phi}^{(1)}(k)+\frac{1}{2}\left(\frac{2\pi}{L}\phi\right)^{2}h_{\phi}^{(2)}(k)$
to the Hamiltonian $H$, where
\begin{align}
h_{\phi}^{(1)}(k) & =\mp\alpha|k\pm k_0|^{\alpha-1},
\label{h1}\\
h_{\phi}^{(2)}(k) & =\mp\alpha\left(\alpha-1\right)\mathrm{sgn}(k\pm k_0)|k\pm k_0|^{\alpha-2},
\label{h2}
\end{align}
with the upper and lower sign corresponding, respectively, to $0<k<\pi$ and $-\pi<k<0$.
The terms of higher orders in $\phi$ do not affect the Thouless conductance defined by
Eq.~(\ref{eq:gTh}). 
We emphasise that the discretisation of momenta, which we use in this paper,
allows us to avoid non-analyticities of $H_{0}(k)$ at the nodal points at $k=\pm \pi/2$.
Utilising Eqs.~\eqref{h1}-\eqref{h2}, we obtain
\begin{align}
\partial_{\phi}^{2}E_{n_0}|_{\phi=0}  =\left(\frac{2\pi}{L}\right)^{2}\left[\left\langle n_0|h_{\phi}^{(2)}\left(k\right)|n_0 \right\rangle
 \right. \nonumber\\ \left.
  +2\sum_{m\neq n_0}\frac{\left|\left\langle m|h_{\phi}^{(1)}\left(k\right)|n_0 \right\rangle \right|^{2}}{E_{n_0}-E_{m}}\right].
  \label{DerivativeDefinition}
\end{align}
In the numerical analysis described below, we use Eq.~\eqref{DerivativeDefinition} in order to determine numerically the Thouless conductance
defined by Eq.~\eqref{eq:gTh}.

%%%%%%%%%%%%%%%%%%%%%%%%%%%%%%%%%%%%%%%%%%%%%%%%%%%%%%%%%%%%%%%%%%%%%%%%%%%%%%%%%%%%%%%%%%%%%%%%%%%%%%%%%%%%%

\section{\label{sec:Thouless}Scaling of Thouless conductance}

\begin{figure}
\includegraphics{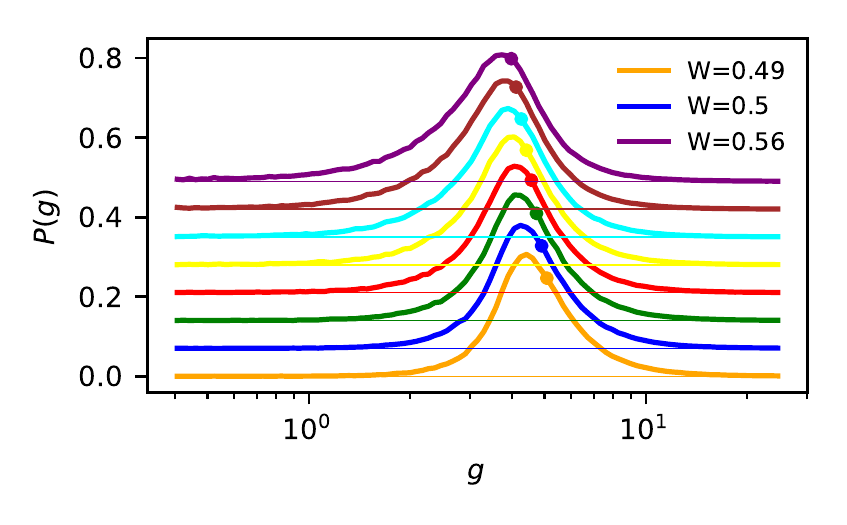}
\caption{\label{fig:Pg}
Normalized probability distribution of the Thouless conductance
obtained for $\sim10^{5}$ disorder realizations in the case $\alpha=0.4$,
 $M=451$ and $W=0.49,0.50,...,0.56$ (bottom to top). The dot indicates the position of the median conductance
$\tilde{g}$. Each curve is offset vertically for better visibility, with the thin horizontal line corresponding to 
vanishing probability density.}
\end{figure}

Near a continous phase transition, the universal critical exponent
$\nu$ describes the divergence of the correlation length~\cite{Cardy:ScalingBook} $\xi\propto |w|^{-\nu}$
as a function of the tuning
parameter that vanishes at the transition point. In this paper, we use the deviation $w=W-W_{c}$ of the disorder amplitude
$W$ from the critical value $W_c$ as the tuning parameter.

The exponent $\nu$ may be determined from the behaviour of a dimensionless observable $\tilde{g}$ that depends on
the system size $L$ only via the dimensionless ratio $L/\xi$, or, equivalently $\tilde{g}(W,L)=\tilde{g}(wL^{1/\nu})$.  
In particular, the critical amplitude $W_{c}$
may be found as the amplitude at which $\tilde{g}(W,L)$ is independent
of the system size $L$. 
% This only holds at zero energy and around critical disorder strength. 
The exponent $\nu$ may then be obtained from a polynomial fitting procedure.~\cite{Slevin:QHPTscaling,Slevin2009:QHPTirrelevantCorrection,Slevin2014:PolynomialFitting}

The outstanding challenge in the context of the non-Anderson transition
is to find an appropriate scaling variable. The density of states
at zero energy, commonly used for the studies of such transitions~\cite{Syzranov:review},
or its dimensionless generalisations, are not suitable 
for finite-size scaling analysis due to the vanishing of the density of states at the transition point
(possibly up small contributions from rare-region effects~\cite{Suslov:rare,Nandkishore:rare,Syzranov:unconv,BuchholdAltland:rare2,Note1}).
Here, we demonstrate numerically that the Thouless conductance
defined by Eq.~(\ref{eq:gTh}) is a suitable scaling variable.

In Fig.~\ref{fig:Pg} we show the probability density of the Thouless conductance for $\alpha=0.4$ and $M=451$,
obtained for $\sim10^{5}$ disorder realizations.
The distribution function $P(g)$ is approximately log-Gaussian.
The scaling hypothesis discussed above applies also to the distribution function $P(g)$.
As it is not practical to work with the full distribution, in what follows
we restrict ourselves to the scaling analysis of only the median $\tilde{g}$.
We note, however, that choosing the typical conductance $\mathrm{exp}(\overline{\mathrm{ln}\,g})$ yields rather similar results.
The standard error of the median conductance is calculated using the asymptotic variance formula $\sigma_{\tilde{g}}^{2}=1/[4P(\tilde{g})^{2}n]$,
where $n$ is the total number of disorder realisations for a particular data point $(W,L)$ and the 
probability density is approximated by a smooth interpolation of the observed histogram.

In Fig. \ref{fig:Summary}a , we plot $\tilde{g}(W,L)$ for $\alpha=0.4$ (corresponding to $|\epsilon|=0.2$) over a range of disorder strengths $W$ and for various system sizes $L$. We observe that data traces for different $L$ cross at one point,
demonstrating the critical scaling behaviour.
To check further the validity of the scaling hypothesis and extract the correlation-length exponent $\nu$, we use the polynomial fitting procedure\cite{Slevin:QHPTscaling,Slevin2009:QHPTirrelevantCorrection,Slevin2014:PolynomialFitting} detailed in Appendix \ref{appendix-FSS}.
This procedure relies on expanding the scaling function $\tilde{g}[u(w)L^{1/\nu}]$ as a low-order polynomial and allows
for quadratic and higher-order corrections in the dependence of the relevant scaling variable $u(w)=w+\cO(w)$ on $w$.
The parameters $\nu$ and $W_{c}$ and the polynomial coefficients are obtained from a least-squares fit (solid lines).

\begin{figure}[h!]
\centering
\includegraphics{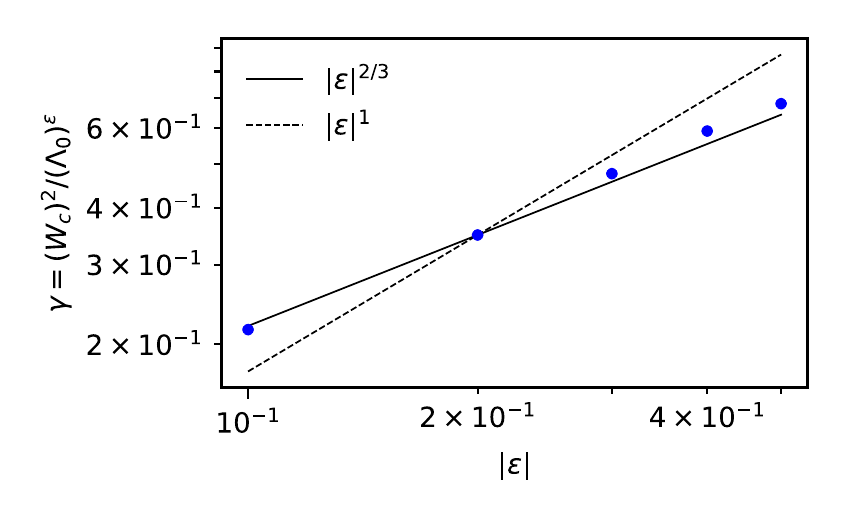}
\caption{\label{fig:gammac}The dimensionless critical disorder strength $\gamma_c=W_{c}^{2}/\Lambda_{0}^{\varepsilon}$
obtained from the finite-size scaling of the Thouless conductance as
a function of $\varepsilon$. The behaviour is consistent with the power-law fitting $\gamma_c\propto|\varepsilon|^{2/3}$ (solid line)
and is distinct from the analytical prediction $\gamma_c\propto|\varepsilon|^{1}$ based on a one-loop perturbative RG calculation
(dashed line).}
\end{figure}

For $\alpha=0.4$ ($|\varepsilon|=0.2$), we find $\nu=\NuFSSpointtwo$.
We use the same method to determine the critical exponent $\nu$
for all values of
$\alpha$ (see Appendix A). The results for $\nu$ and $W_c$ as a function of $\varepsilon=2\alpha-1$ are shown,
respectively, in Figs.~\ref{fig:Summary}b and \ref{fig:gammac}.

%%%%%%%%%%%%%%%%%%%%%%%%%%%%%%%%%%%%%%%%%%%%%%%%%%%%%%%%%%%%%%%%%%%%%%%%%%%%%%%%%%%%%%%%%%
%%%%%%%%%%%%%%%%%%%%%%%%%%%%%%%%%%%%%%%%%%%%%%%%%%%%%%%%%%%%%%%%%%%%%%%%%%%%%%%%%%%%%%%%%%

\section{\label{sec:DOS}Scaling of the density of states}

Most studies to date have relied on the scaling form~\cite{KobayashiOhtsukiHerbut:scaling}
of the density of states for the numerical determination
of the critical exponents.
%The earliest attempt to numerically determine the critical exponent
%$\nu$ in the non-Anderson transition for a three-dimensional Dirac
%node employed the density of states $\rho(E=0)$ as a scaling variable \cite{KobayashiOhtsukiHerbut:scaling}.
In what immediately follows, we demonstrate that
the scaling of the density of states in the model considered here is consistent with the results
obtained from the finite-size scaling analysis of the Thouless conductance.

\begin{figure}[h!]
\centering
\includegraphics{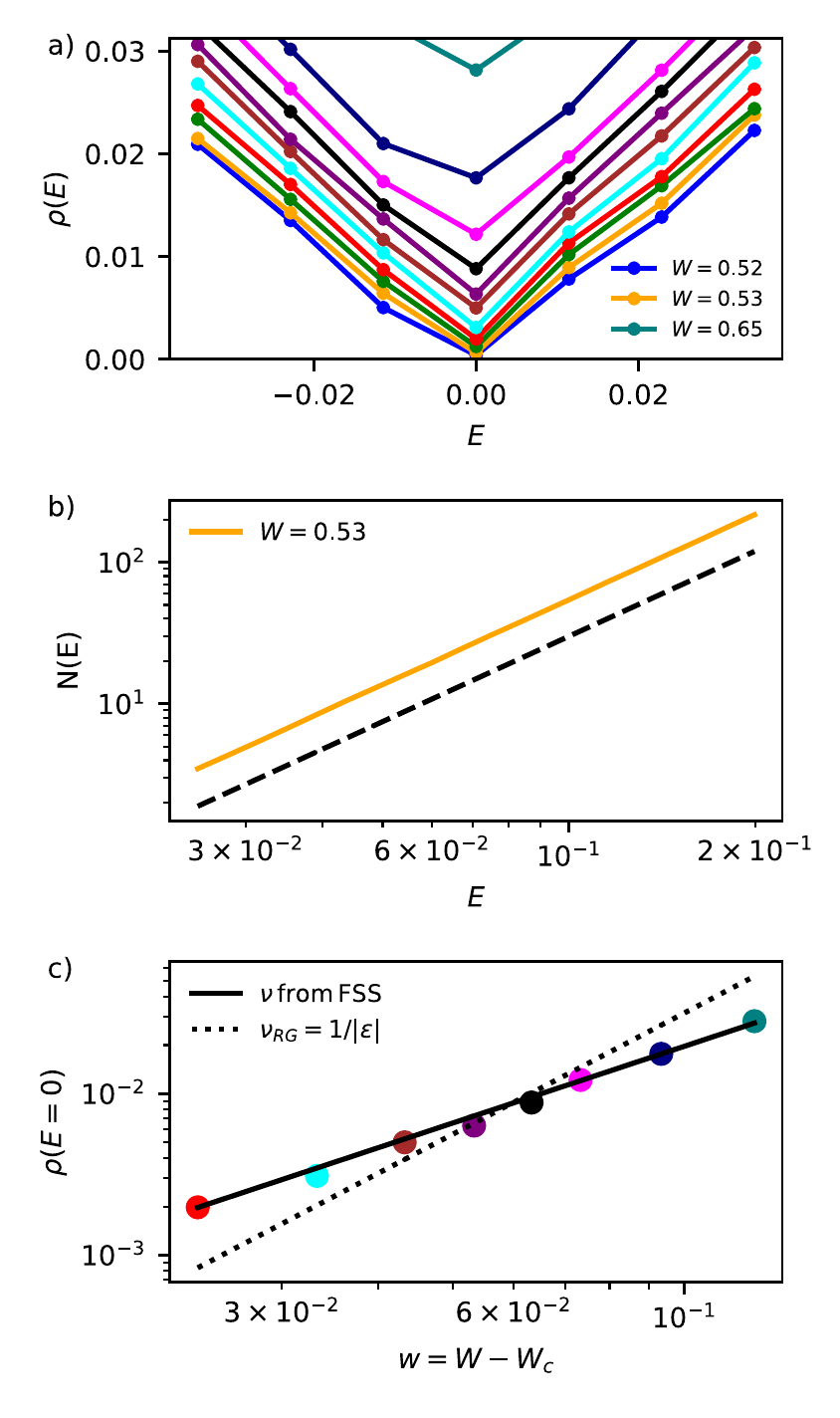}
\caption{\label{fig:DOS}
	(a) Density of states $\rho(E,W)$ for $\alpha=0.4$
and $M=8001$ (dots). The solid lines are guides to the eye. (b) The
integrated density of states close to criticality ($W=0.53$) is consistent
with the scaling form $N(E)\propto E^{1/z}$ with $z=1/2$ (dashed line).
(c) The zero-energy density of states (dots) follows the
scaling $\rho(E=0,W>W_{c})\propto w^{(1-z)\nu}$ with $z=1/2$ and
$\nu=3.163$ obtained from the finite-size scaling
(solid line), but not with the analytical RG prediction $\nu_{RG}=1/|\varepsilon|$ (dotted line).}
\end{figure}

As discussed in Sec.~\ref{sec:Thouless}, simulating numerically the density of states
does not allow for an accurate determination of the critical disorder amplitude $W_c$,
which is why we use the result $W_{c}=0.5267$ obtained from the finite-size scaling of the Thouless conductance
to fit the scaling of the density of states for a system with $\alpha=0.4$. 
The disorder-averaged density of states
$\rho(E\approx 0)$ for the system size $M=8001$ and various
disorder strengths around $W_{c}$ is shown in Fig. \ref{fig:DOS}a.

%{I stopped here. 24 Feb}

We obtain that the density of states $\rho(E\approx 0,W)$ is strongly suppressed for
$W \leq W_c$ and grows rapidly at larger $W$. Our results are consistent, to a good accuracy,
with the scaling form~\cite{KobayashiOhtsukiHerbut:scaling}
\begin{equation}
\rho(E=0,W>W_{c})\propto (W-W_c)^{(1-z)\nu}\label{eq:DOS_E=0},
\end{equation}
%In the one-dimensional case, the scaling form for the density of states
%at zero energy reads \cite{KobayashiOhtsukiHerbut:scaling, Garttner:longrange}
where $z$ is the dynamical critical exponent, given
by $z=1/2+\cO(\varepsilon)$ according to the perturbative RG calculations.
Numerically, the value of the dynamical exponent may be obtained from the scaling of the integrated
density of states (i.e. the number of states with energies between $0$ and $E$) 
$N(E)=\int_{0}^{E}d\epsilon\,\rho(\epsilon)\propto E^{1/z}$ at $W=W_c$.
Our data for $W=0.53$ close
to $W_{c}$ is consistent with $z=1/2$ (dashed line) within a few-percent error.

In Fig. \ref{fig:DOS}c, we show the numerical data for $\rho(E=0,W>W_{c})$
vs. $W-W_{c}$ in a log-log plot (dots). 
The dependence is consistent with Eq.~(\ref{eq:DOS_E=0}) with the correlation-length exponent $\nu=3.163$
obtained from the finite-size scaling above (solid line).
On the contrary, the prediction $\nu_{RG}=1/|\varepsilon|$ of the one-loop RG (dashed
line) is clearly inconsistent with the numerical data.

%%%%%%%%%%%%%%%%%%%%%%%%%%%%%%%%%%%%%%%%%%%%%%%%%%%%%%%%%%%%%%%%%%%%%%%%%%%%%%%%%%%%%%%%%%%%%%%%%%%%%%%%%%%%%%%%%%%%%%%%%%%%%%%%
%%%%%%%%%%%%%%%%%%%%%%%%%%%%%%%%%%%%%%%%%%%%%%%%%%%%%%%%%%%%%%%%%%%%%%%%%%%%%%%%%%%%%%%%%%%%%%%%%%%%%%%%%%%%%%%%%%%%%%%%%%%%%%%%

\section{\label{sec:Discussion}Discussion and outlook}

%The analysis of conventional field theories showing quantum phase transitions is notoriously difficult. For analytical approaches, one often singles out parameters like the inverse number of particle flavors $1/N$ or the distance to some critical dimension $\varepsilon$ and treats them as small, even though physically they are of oder unity. In terms of numerics, system sizes are often severley limited by the interacting nature of the system in question. 

To summarise, in this paper we proposed and studied a one-dimensional model for the non-Anderson disorder-driven transitions.
This model presents a rather convenient platform for investigating the non-Anderson criticality, as the system
has a tunable small parameter $\varepsilon$ that controls the criticality.
The model considered here
exhibits the critical scaling of an observable (Thouless conductance) which does not vanish at the transition,
unlike the density of states usually used in such studies, and thus allows for an accurate characterisation of the criticality.
In our model, the energies of the states, for which the non-Anderson transition occurs, are not subject to renormalisation by disorder,
which allows us to additionally increase the accuracy of the results. 

Utilising the small parameter $\varepsilon$ allows us not only to study numerically the non-Anderson criticality within the expected range
of applicability of the analytical perturbative RG approaches, but also to neglect possible non-perturbative rare-region
effect~\cite{Suslov:rare,Nandkishore:rare,Note1}, exponentially suppressed by the large parameter $1/|\varepsilon|$ in the limit $\varepsilon\rightarrow0$. 
This ensures the observability of the critical scaling in a parametrically large interval of disorder strengths and length scales.

Furthermore, the approach we develop here may be used to analyse
the convergence of $\varepsilon$-expansions in interacting field theories similar to those describing 
the disordered high-dimensional systems.

Our data demonstrates the correlation-length critical exponent $1/\nu=3|\varepsilon|/2+\cO(\varepsilon^2)$ in the limit of small
$\varepsilon$, which, contrary to the common expectation, is
distinct from the prediction $1/\nu_{RG}=|\varepsilon|+\cO(\varepsilon^2)$ of the perturbative RG approach.
Furthermore, the scaling of the dependence of the critical disorder strength on $|\varepsilon|$
is given by $\gamma_{c}\propto |\varepsilon|^\frac{2}{3}$, in accordance with the earlier study in Ref. \onlinecite{Moura:firstPowerLaw}
and in disagreement with the expectation $\gamma_{c}\propto |\varepsilon|$ from the perturbative RG analysis.
At the same time, the numerical result for dynamical exponent, $z=\frac{1}{2}+\cO(\varepsilon)$, agrees well with RG predictions.
Similar agreement between the numerical values of the dynamical exponent $z$ and the one-loop RG results are observed
for Weyl semimetals, while the results for the correlation-length exponent $\nu$ are controversial (see Ref.~\onlinecite{Syzranov:review}
for a review). 

We leave for future studies the origin of the disagreement
between the numerical and the perturbative RG results for the correlation-length exponent in the limit of small $\varepsilon$.
This disagreement may mean the breakdown of small-$\varepsilon$ expansions in the RG approaches or, for example, the existence of
spontaneously generated operators in the field theories for non-Anderson transitions, which have been overlooked
in the previous analytical studies of these transitions.

Such a possibility and the controversy around the values of the correlation-length exponents in Weyl semimetals
call for a fresh analytical investigation of the non-Anderson criticality in these systems.

Assuming that the critical behaviour at the criticality is determined by one relevant coupling $\tilde{\gamma}$,
our numerical results for the dependence of the critical disorder strength and the correlation-length exponent on $\varepsilon$
are consistent with the beta-function $\beta(\tilde\gamma)=\varepsilon\tilde\gamma+(\tilde{\gamma})^\frac{5}{3}+\ldots$,
where $\ldots$ stands for the terms of higher orders in $\tilde{\gamma}$. We leave for future studies the investigation of the possible nature of 
such variables.

%%%%%%%%%%%%%%%%%%%%%%%%%%%%%%%%%%%%%%%%%%%%%%%%%%%%%%%%%%%%%%%%%%%%%%%%%

\section{Acknowledgements.} We thank Thomas Klein-Kvorning and Vlad Kozii for useful discussions. Our work has been financially supported by the German National Academy of Sciences Leopoldina through grant LPDS 2018-12 (BS)
and the Hellman foundation (SS).

%%%%%%%%%%%%%%%%%%%%%%%%%%%%%%%%%%%%%%%%%%%%%%%%%%%%%%%%%%%%%%%%%%%%%%%%%%%%%%%%%%%%%%%%%%%%%%%%%%%%%
%%%%%%%%%%%%%%%%%%Appendix%%%%%%%%%%%%%%%%%%%%%%%%%%%%%%%%%%%%%%%%%%%%%%%%%%%%%%%%%%%%%%%%%%%%%%%%%%%

\onecolumngrid
%\vspace{2cm}
%\twocolumngrid

\newpage

\appendix

\section{\label{appendix-FSS}Details of finite-size scaling and additional data}
\begin{figure}[H]
	\centering
	\includegraphics[scale=0.9]{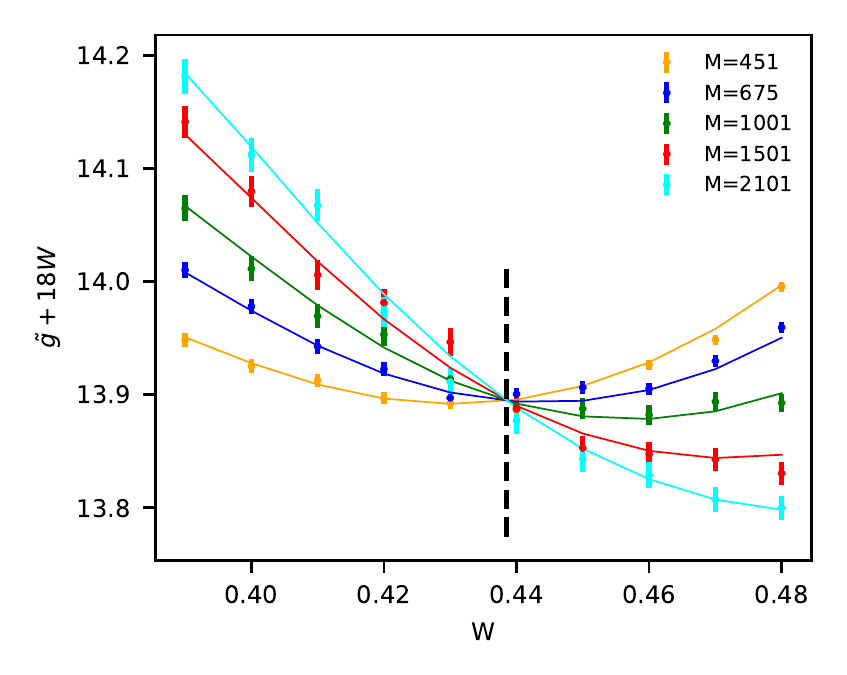}
	\includegraphics[scale=0.9]{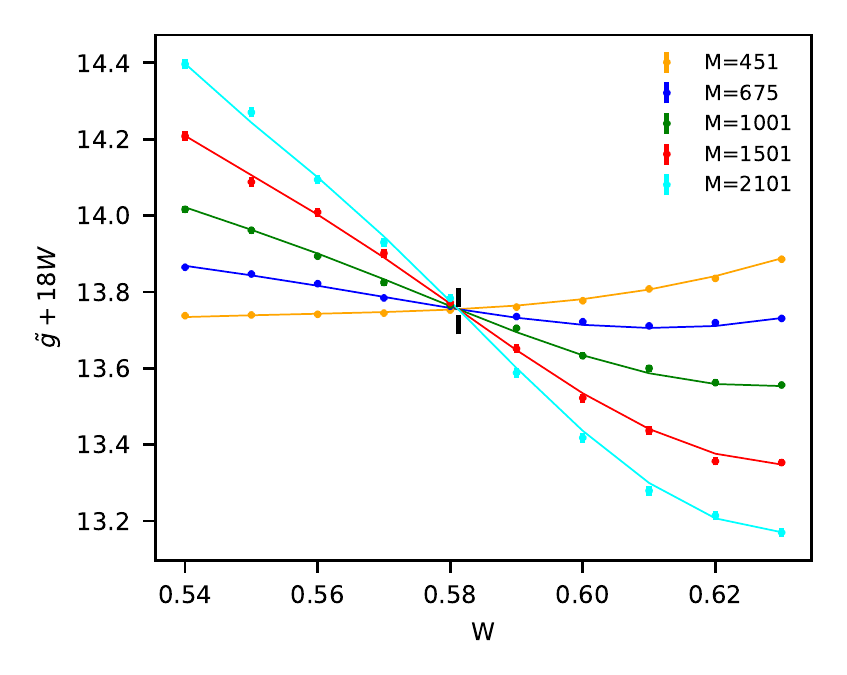}
	\vspace{10mm}
	\includegraphics[scale=0.9]{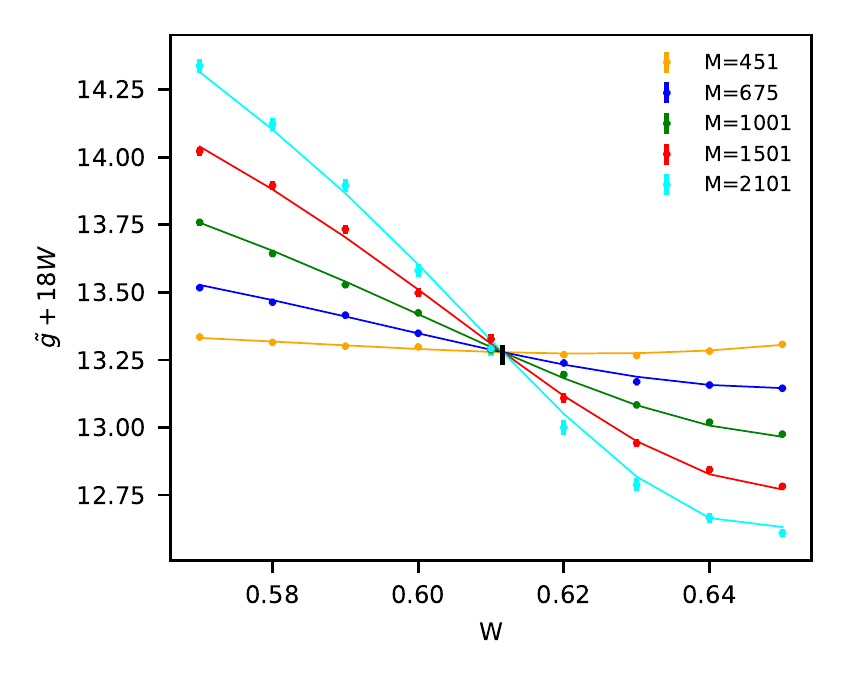}
	\includegraphics[scale=0.9]{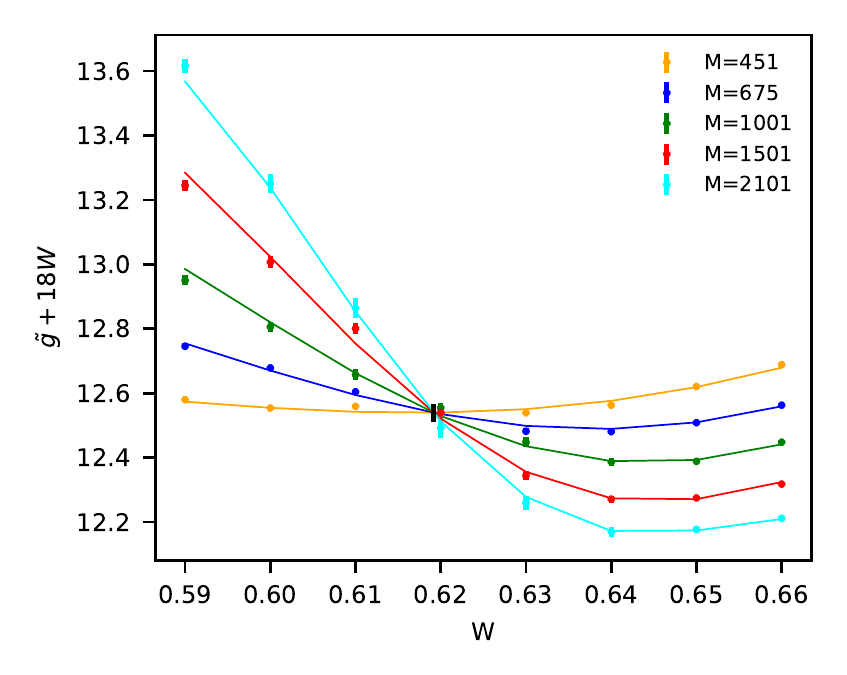}
	
	\caption{\label{fig:otherAlphaFSS}Finite-size scaling data as in Fig. \ref{fig:Summary}a,
		but for $\alpha=0.45$, $0.35$, $0.3$, $0.25$ (left to right and top to bottom).
		%\tg{For some reason, all of these figures appear in the very end. Can we do anything to make them show up where we want them?}
	}
\end{figure}

Here, we provide the details of the finite-size scaling data analysis, following
Refs. \onlinecite{Slevin:QHPTscaling,Slevin2009:QHPTirrelevantCorrection,Slevin2014:PolynomialFitting}. We consider the cases of power-law
exponent $\alpha=\{0.45,0.4,0.35,0.3,0.25\}$, each for system sizes
$L=2M$ with $M=\{451,675,1001,1501,2101\}$ and a range of disorder
strengths $W$ in the vicinity of $W_{c}$, summarised in Table \ref{tab:FSS}.
For each $\alpha$ and each point $(W,L)$, we determine the median
$\tilde{g}(W,L)$ of the conductance distribution and its uncertainty,
$\sigma_{\tilde{g}}(W,L)$. We use between $\sim10^{4}$ and $\sim10^{5}$
disorder realizations, depending on the system size. The data is shown
as symbols in Figs. \ref{fig:Summary}b,c and \ref{fig:otherAlphaFSS}. We then perform a least-squares fit
\begin{equation}
\tilde{g}_{\mathrm{fit}}(W,L)=a_{0}+\sum_{p=1}^{n_{p}}a_{p}\cdot\left(L^{1/\nu}\sum_{q=1}^{n_{q}}b_{q}\left[W-W_{c}\right]^{q}\right)^{p},\label{eq:fit}
\end{equation}
with $b_1=1$ for various orders $n_p$ and $n_q$ and obtain the quality of the fit in terms
of 
\begin{equation}
\chi^{2}/N=\frac{1}{N}\sum_{(W,L)}\frac{\left[\tilde{g}_{\mathrm{fit}}(W,L)-\tilde{g}(W,L)\right]^{2}}{\sigma_{\tilde{g}}^{2}(W,L)}\label{eq:chi2oN}
\end{equation}
where $N$ is the total number of data points $(W,L)$ used in the
fit. We increase $n_p$ and $n_q$ if this lowers the value of $\chi^{2}/N$ by
more than 2\% and the error of any fitting parameter does not exceed
the parameter estimate in magnitude. We find that $n_p=4$ and $n_q=2$ yield the
optimal fit in this sense. 
We verify also that the inclusion of possible irrelevant scaling variables~\cite{Slevin:QHPTscaling}, corresponding to the replacement 
$a_{i}\rightarrow a_{i}\left(1+c_{i}L^{y}\right)$ with $y<0$ in Eq. (\ref{eq:fit}), does not affect significantly the value of $\chi^{2}/N$
and does not lead to stable values of $y$, which allows us to neglect irrelevant scaling variables.

Each fitting procedure is repeated a hundred times with randomly chosen
initial parameters. For the best fits with the lowest $\chi^{2}/N$, Table \ref{tab:FSS} reports the resulting
parameters $\nu$, $W_{c}$, their error estimates and $\chi^{2}/N$. The best fits are shown by solid lines in Figs.~\ref{fig:Summary}a and \ref{fig:otherAlphaFSS}.
We find that removing the largest or smallest length data set ($M=451$
or $M=2101$) from the fitting procedure does produce consistent results
with the estimates for $\nu$ and $W_{c}$ within
the previous error bars (data not shown).

\noindent \begin{center}

\begin{table}[H]
\noindent \begin{centering}
\begin{tabular}{|c|c|c|c|c|c|c|c|}
\hline 
$\alpha$ & $|\epsilon|$ &$W$ & $M=L/2a$ & $N$ & $\chi^{2}/N$ & $\nu$ & $W_{c}$\tabularnewline
\hline 
\hline 
0.45 &0.1& 0.39,0.40,...,0.48 & 451,675,1001,1501,2101 & 50 & 0.920 & 6.196$\pm$0.171 & 0.4385$\pm$0.0010\tabularnewline
\hline 
0.40 &0.2& 0.49,0.495...,0.56 & 451,675,1001,1501,2101 & 75 & 0.797 & \NuFSSpointtwo  & 0.5267$\pm$0.0004\tabularnewline
\hline 
0.35 &0.3& 0.54,0.55,...,0.63 & 451,675,1001,1501,2101 & 50 & 0.915 & 2.111$\pm$0.021 & 0.5812$\pm$0.0003\tabularnewline
\hline 
0.30 &0.4& 0.57,0.58,...,0.65 & 451,675,1001,1501,2101 & 45 & 1.277 & 1.719$\pm$0.017 & 0.6116$\pm$0.0003\tabularnewline
\hline 
0.25 &0.5& 0.59,0.60,...,0.66 & 451,675,1001,1501,2101 & 40 & 1.588 & 1.532$\pm$0.022 & 0.6191$\pm$0.0003\tabularnewline
\hline 
\end{tabular}
\par\end{centering}
\caption{\label{tab:FSS}Details for the finite-size scaling: The left two columns
specify the value of $\epsilon=2\alpha-1$, the third and fourth the
disorder strength $W$ and system size $L=2M$, respectively. The total
number of data points $(W,L)$ is denoted by $N$. The right part of the table
reports the $\chi^{2}/N$ value for the best fit along with its estimate
for the critical exponent $\nu$ and the critical disorder strength
$W_{c}$. The error (denoted after the $\pm$ symbol) is one standard
deviation. The respective plots can be found in Fig. \ref{fig:Summary}a ($\alpha=0.4$)
and Fig. \ref{fig:otherAlphaFSS} ($\alpha=0.45$, $0.35$, $0.3$,
$0.25$).}
\end{table}
\par\end{center}

%%%%%%%%%%%%%%%%%%%%%%%%%%%%%%%%%%%%%%%%%%%%%%%%%%%%%%%%%%%%%%%%%%%%%%%%%%%%%%%%%%%%%%%%%%%%%%%%%%%%%%%%%%%%%%%%%
%%%%%%%%%%%%%%%%%%%%%%%%%%%%%%%%%%%%%%%%%%%%%%%%%%%%%%%%%%%%%%%%%%%%%%%%%%%%%%%%%%%%%%%%%%%%%%%%%%%%%%%%%%%%%%%%%

\section{\label{appendix-RG}Perturbative renormalisation-group analysis}

\begin{figure}[h!]
	\centering
	\includegraphics[width=0.45\linewidth]{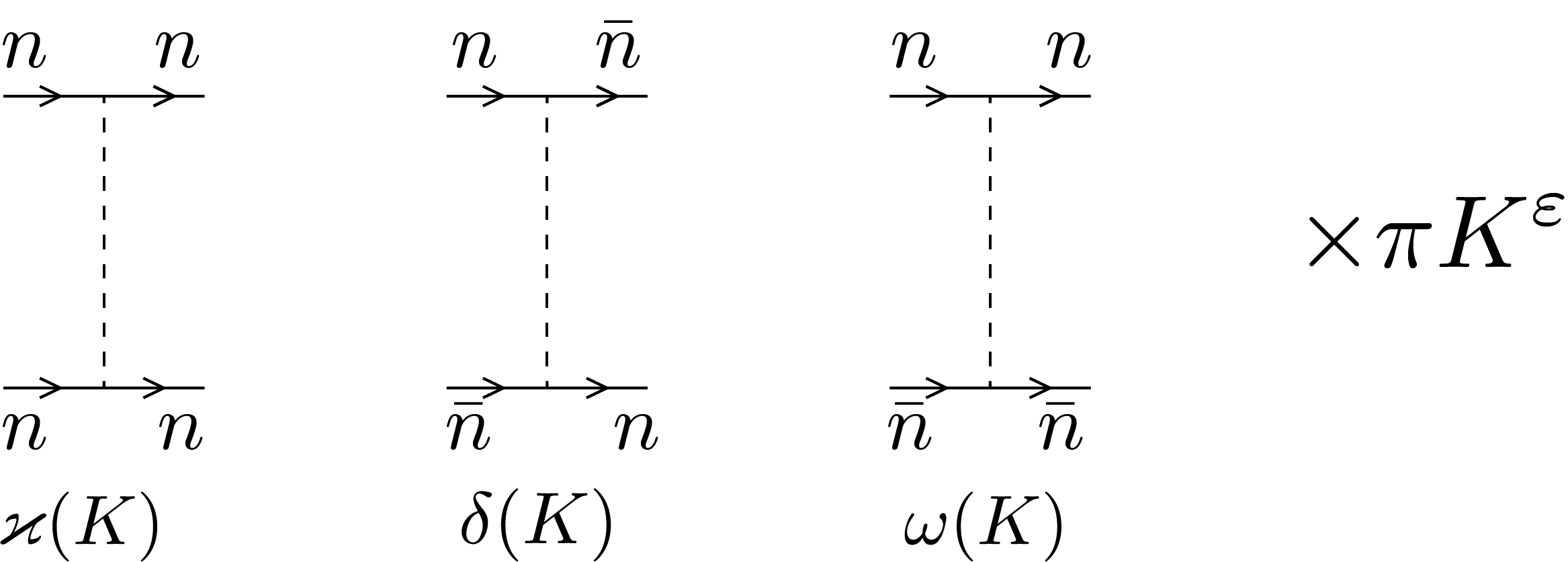}
	\caption{\label{fig:implines} 
		Elements of the diagrammatic technique (impurity lines) used to compute disorder-averaged observables in a model with two nodes
		described by the Hamiltonian~\eqref{eq:H0}.
		The coupling constants $\varkappa(K)$, $\delta(K)$ and $\omega(K)$ depend on the
	ultraviolet momentum cutoff $K$. }
\end{figure}

In this section, we provide a detailed one-loop RG analysis for a model with the two-node quasiparticle 
dispersion~\eqref{eq:H0} in the presence of random potential with a short correlation length.
%The random potential is assumed to be correlated on length scales significantly shorter than the inverse separation between
%the nodes in momentum space. 
The wavevector $k$ of the quasiparticles we consider is measured from either of the nodes $n=1,2$ and is exceeded significantly 
by the inverse characteristic correlation length of the random potential.
Disorder-averaged observables in this systems may be described by a supersymmetric~\cite{Efetov:book} field theory with the action
\begin{align}
	\cL=
	&-i\sum\limits_{n=1,2}\left\{
	\int \bar\psi_n \left[ - k^\alpha \sign k + i0\Lambda \right]\psi_n\, dx
	\right.
	\nonumber\\
	&\left.+\frac{1}{2}\varkappa(K)K^{\varepsilon}\int \left(\bar{\psi}_n\psi_n\right)^2 dx
		+\frac{1}{2}\delta(K)K^{\varepsilon}\int \left(\bar{\psi}_n\psi_{\bar{n}}\right)\left(\bar{\psi}_{\bar{n}}\psi_n\right) dx
		+\frac{1}{2}\omega(K)K^{\varepsilon}\int \left(\bar{\psi}_n\psi_n\right)\left(\bar{\psi}_{\bar{n}}\psi_{\bar{n}}\right) dx
	\right\},
	\label{Action}
\end{align}
where $\bar\psi_n(x)$ and $\psi_n(x)$ are the supervectors in the $BF\otimes AR$ (boson-fermion$\otimes$advanced-retarded) space describing the quasiparticles near node $n$; 
$\bar{n}$ is our convention for 
the node other than $n$, i.e. $\bar{1}=2$ and $\bar{2}=1$; $\Lambda=\left(\hsigma_z\right)_{AR}$; $k=-i\partial_x$ is the momentum operator;
 $K$ is the ultraviolet momentum cutoff.
The coupling constants $\varkappa(K)$, $\delta(K)$ and $\omega(K)$ which characterise scattering processes shown in Fig.~\ref{fig:implines}
flow under renormalisation.

\begin{figure}%[H]
	\centering
	\includegraphics[width=0.9\linewidth]{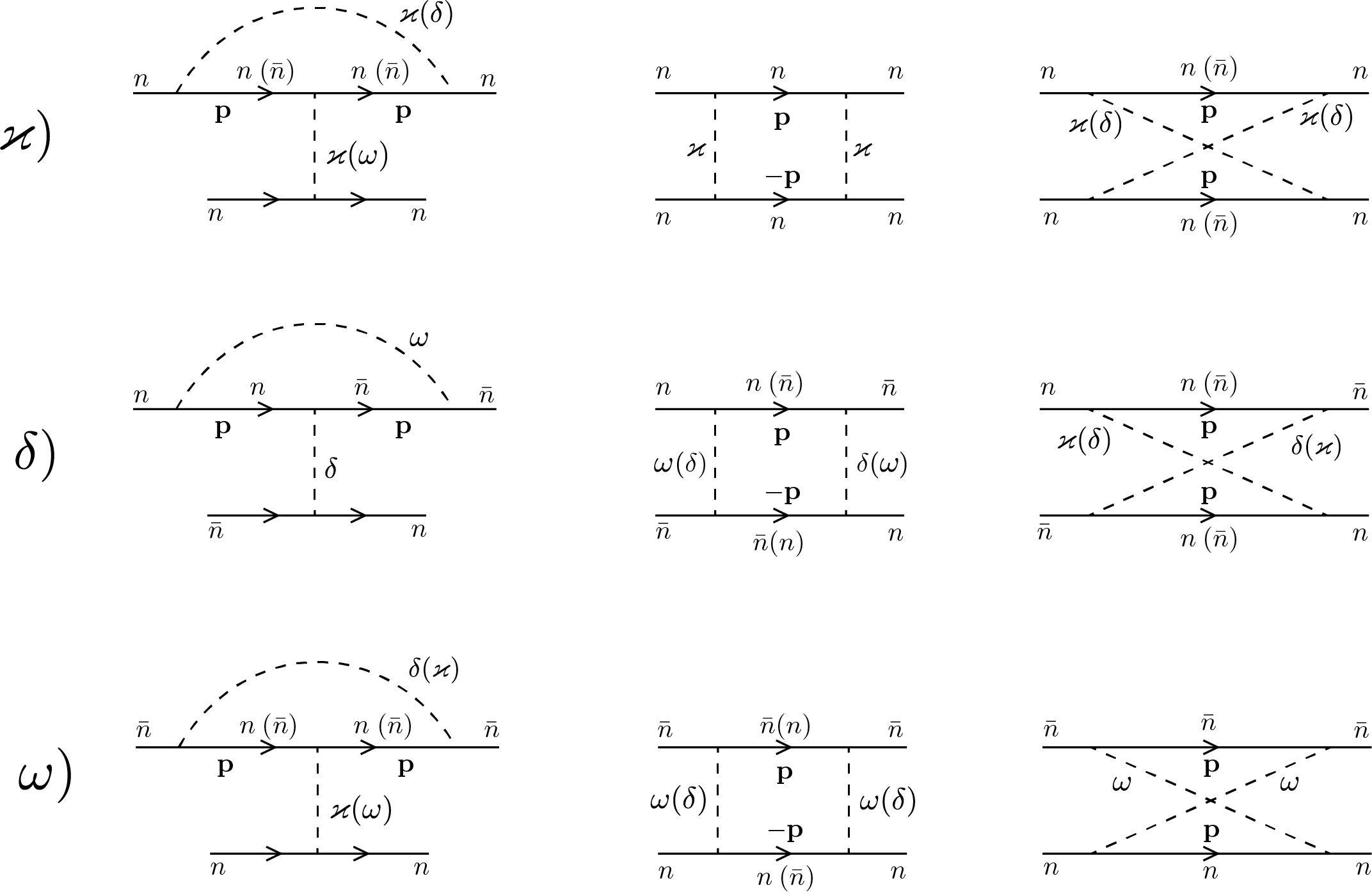}
	\caption{\label{fig:rgdiagrams} Diagrams (in units of $\pi^2 K^{2\varepsilon}$) contributing to the one-loop renormalisations of the coupling constants $\varkappa$, $\delta$ and $\omega$. 
	Symbols in parentheses indicate alternative nodes or types of impurity lines.
	For each shown ``mushroom'' diagram, there exists one equivalent diagram with an impurity line connecting two points on the bottom 
	particle propagator in place of the top propagator.}	
\end{figure}

The renormalisation procedure involves {repeatedly integrating out} particle modes with highest momenta, starting at the ultraviolet momentum
cutoff $K=K_0$ and correcting the action of the low-momentum modes to account for the effect of the higher momenta.
Upon renormalisation, the action \eqref{Action} reproduces itself with flowing couplings $\varkappa(K)$, $\delta(K)$
and $\omega(K)$. The perturbative one-loop corrections to each of the three coupling constants are shown in Fig.~\ref{fig:rgdiagrams}.
The momenta, which we integrate out, are away from the poles of the Green's functions and, therefore, both advanced and 
retarded Green's functions may be approximated as
$G_{1,2}(p)\approx \mp\frac{\sign p}{p^\alpha}$.

The flows of the coupling constants $\varkappa(K)$, $\delta(K)$ and $\omega(K)$ may be represented in the form
\begin{subequations}
\begin{align}
	\partial_l \varkappa = &\beta_\varkappa (\varkappa,\delta,\omega), \label{VarkappaBeta}\\	
	\partial_l \delta =	& \beta_\delta (\varkappa,\delta,\omega),\\
	\partial_l \omega =	& \beta_\omega (\varkappa,\delta,\omega),
\end{align}
\end{subequations}
where the beta-functions 
\begin{subequations}
	\begin{align}
	\beta_\varkappa (\varkappa,\delta,\omega) =& \varepsilon\varkappa +2\varkappa^2 +2\omega\delta+\delta^2 +\ldots,
	\label{VarkappaFlow} \\
	\beta_\delta (\varkappa,\delta,\omega) =& \varepsilon\delta+ 2\delta\varkappa +\ldots, 
	\label{DeltaFlow} \\
	\beta_\omega (\varkappa,\delta,\omega) =& \varepsilon\omega + 2\delta\varkappa + 2\varkappa\omega+ \delta^2 +\ldots.
	\label{OmegaFlow}
	\end{align}
\end{subequations}
are given, to the one-loop order, by the sum of the diagrams in Fig.~\eqref{fig:rgdiagrams} and $\ldots$ are higher-order terms in the coupling constants.

%Summing up the contributions shown in Fig.~\ref{fig:rgdiagrams} gives that the coupling constants flow (in the one-loop approximation) as 
%\begin{subequations}
%	\begin{align}
%	\partial_{l}\varkappa =& \varepsilon\varkappa +2\varkappa^2 +2\omega\delta+\delta^2,
%	\label{VarkappaFlow} \\
%	\partial_{l}\delta =& \varepsilon\delta + 2\delta\varkappa, 
%	\label{DeltaFlow} \\
%	\partial_{l}\omega =& \varepsilon\omega + 2\delta\varkappa + 2\varkappa\omega + \delta^2.
%	\label{OmegaFlow}
%	\end{align}
%\end{subequations}
%In this paper, we consider short-range correlated disorder, which corresponds to matching coupling constants $\varkappa(K_0)=\omega(K_0)$
%at the initial ultraviolet cutoff $K_0$. Under such initial conditions, these coupling constants match at all values of the ultraviolet cutoff,
%$\varkappa(K)=\omega(K)$, according to Eqs.~\eqref{VarkappaFlow}-\eqref{OmegaFlow}.

The fixed points of the flow are determined by the conditions $\beta_\varkappa (\varkappa^*,\delta^*,\omega^*)=\beta_\delta (\varkappa^*,\delta^*,\omega^*)
=\beta_\omega (\varkappa^*,\delta^*,\omega^*)=0$.
Utilising Eq.~\eqref{DeltaFlow}, this requires that either $\delta^*=-\frac{\varepsilon}{2}$ or $\delta^*=0$. Using
that $\varkappa(K)=\omega(K)$ for the 
initial conditions for the flow under consideration and the form of the beta-functions $\beta_\delta$
and $\beta_\omega$ given by Eqs.~\eqref{VarkappaFlow} and \eqref{OmegaFlow},
we obtain the values of the coupling constants at two fixed points (to the leading order in $|\varepsilon|$):
1) $\delta^*=0$, $\varkappa^*=\omega^*=-\frac{\varepsilon}{2}$ and
2) $\delta^*=\varepsilon$, $\varkappa^*=\omega^*=-\frac{\varepsilon}{2}$.
We note that for the model under consideration all couplings remain positive under renormalisation and, therefore, the second
fixed point is unachievable as it corresponds to a negative coupling $\delta$.

The correlation-length exponents $\nu$ at the transition are given by the inverse eigenvalues of the
Jacobian $\frac{\partial\left(\beta_\varkappa,\beta_\delta,\beta_\omega\right) }{\partial\left(\varkappa,\delta,\omega\right)}$.
Utilising Eqs.~\eqref{VarkappaFlow}-\eqref{OmegaFlow}, {we obtain at the first fixed point
\begin{align}
	\frac{\partial\left(\beta_\varkappa,\beta_\delta,\beta_\omega\right) }{\partial\left(\varkappa,\delta,\omega\right)}
	(\varkappa^*,\delta^*,\omega^*)
	=
	\left(
	\begin{array}{ccc}
	-\varepsilon & -\varepsilon & 0 \\
	0 & 0 & 0 \\
	-\varepsilon & -\varepsilon & 0
	\end{array}
	\right)+\cO\left(\varepsilon^2\right).
	\label{Jacobian}
\end{align}
The Jacobian~\eqref{Jacobian} gives the correlation-length exponent $1/\nu_{RG}=-\varepsilon$.}

\begin{figure}%[H]
	\centering
	\includegraphics[width=0.4\linewidth]{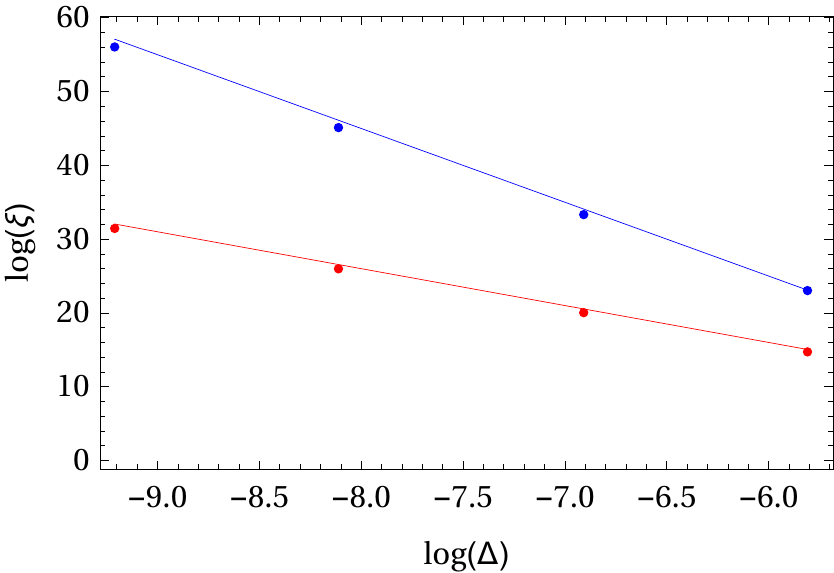}
	\caption{\label{fig:numericalRG} Verification of the scaling form $\xi \propto \Delta^{-\nu}$ for $\Delta = \gamma_0-\gamma_c$ based on the numerical solution of the flow equations (\ref{VarkappaBeta}) -- (\ref{OmegaFlow}). The numerically extracted correlation length data (dots) for $\varepsilon=-0.1$ (blue) and $\varepsilon=-0.2$ (red) lie on straigth lines with the slope $-\nu_{RG}=-1/|\varepsilon|$.}	
\end{figure}

In addition to the above analysis of the fixed points, we verify {the correlation-lenght exponent} by solving the RG equations
 (\ref{VarkappaBeta}) -- (\ref{OmegaFlow}) numerically with the initial conditions $\kappa(K_0)=\delta(K_0)=\omega(K_0)=\gamma_0$ 
corresponding to the short-range-correlated random potential considered in this paper.
First, we identify the critical disorder strength $\gamma_c$ by varying $\gamma$ for fixed $\varepsilon<0$
and detecting the value of $\gamma$ at which the flows become unstable.
Next, we determine the correlation-length exponent by solving numerically the RG equations for 
the supercritical disorder strength $\gamma_0=\gamma_c+\Delta$ with a small $\Delta>0$, where we vary $\Delta$ such that it is an order of magnitude larger than the uncertainty of $\gamma_c$ and an order of magnitude smaller than $\gamma_c$. We find the RG time $l=\tilde{l}$ where the flow of $\kappa(l)=\omega(l)$ diverges and the numerical solution breaks down. Due to the extreme sharpness of the divergence,
 this agrees with the more common requirement $\kappa(\tilde{l})=\omega(\tilde{l})=C$ with $C=\mathcal{O}(1)$. The RG time $\tilde{l}$ is related to the emergent correlation length $\xi$ as $\tilde{l} \sim \mathrm{log}\xi $. In Fig. \ref{fig:numericalRG}, we confirm the scaling form $\xi \propto \Delta^{-\nu}$ by plotting $\mathrm{log}(\xi)$ versus $\mathrm{log}(\Delta)$. Indeed, the data points for $\varepsilon=-0.1$ (blue) and $\varepsilon=-0.2$ (red) lie on straight lines with slopes $-\nu_{RG}=-1/|\varepsilon|$, confirming the results from the fixed-point analysis in the previous paragraph. In addition, we confirm also that $\gamma_c \propto |\varepsilon|$.

\FloatBarrier

%%%%%%%%%%%%%%%%%%%%%%%%%%%%%%%%%%%%%%%%%%%%%%%%%%%%%%%%%%%%%%%%%%%%%%%%%%%%%%%%%%%%%%%%%%%%%%%%%%%%
%%%%%%%%%%%%%Bibliography%%%%%%%%%%%%%%%%%%%%%%%%%%%%%%%%%%%%%%%%%%%%%%%%%%%%%%%%%%%%%%%%%%%%%%%%%%%

\twocolumngrid

\bibliography{references}

%%%%%%%%%%%%%%%%%%%%%%%%%%%%%%%%%%%%%%%%%%%%%%%%%%%%%%%%%%%%%%%%%%%%%%%%%%%%%%%%%%%%%%%%%%%%%%%%%%%%
%%%%%%%%%%%%%%%%%%%%%%%%%%%%%%%%%%%%%%%%%%%%%%%%%%%%%%%%%%%%%%%%%%%%%%%%%%%%%%%%%%%%%%%%%%%%%%%%%%%%

\end{document}